\documentclass[letterpaper, 10 pt, conference]{ieeeconf}
    
    \IEEEoverridecommandlockouts 
    \makeatletter

    \let\proof\@undefined
    \let\endproof\@undefined
    \makeatother
    \let\labelindent\relax 

    \usepackage{jhoagg}
    \usepackage{mathtools}
    \usepackage{amsfonts}
    \usepackage{amssymb,latexsym}
    \usepackage[psamsfonts]{eucal}
    \usepackage{amsthm}
    \usepackage{graphicx}
    \usepackage[inline]{enumitem}
    \usepackage{indentfirst}
    \usepackage{setspace}
    \usepackage{microtype}
    \usepackage{threeparttable}
    \usepackage{float}
    \usepackage{cleveref}
    \usepackage{afterpage}
    \usepackage{placeins}
    \usepackage{cancel}
    \usepackage{leftidx}
    \usepackage{breqn}
    \usepackage[export]{adjustbox}
    \usepackage{comment}
    \usepackage[ruled, linesnumbered, lined]{algorithm2e}
    \usepackage[dvipsnames]{xcolor}
    \usepackage{xcolor}
    \usepackage[most]{tcolorbox}
    \usepackage{soul}
    \usepackage[noadjust]{cite}
    \usepackage{standalone}
    \usepackage{dsfont}

    \usepackage{tikz}
    \usetikzlibrary{shapes,arrows,positioning,fit}

    \usepackage{pgf}
    \usepackage{pgfplots}
    \usepgflibrary{plothandlers}
    \usepgfplotslibrary{external} 
    \pgfkeys{/pgf/number format/.cd,1000 sep={}}  
    \usetikzlibrary{calc,shapes.misc,plotmarks}
    \pgfplotsset{compat=1.13}
    
    \let\originalleft\left
    \let\originalright\right
    \renewcommand{\left}{\mathopen{}\mathclose\bgroup\originalleft}
    \renewcommand{\right}{\aftergroup\egroup\originalright}
    
    \usepackage{fancyhdr}
    \fancypagestyle{firststyle}
    {\fancyfoot{} 
    \fancyfoot[L]{\textit{Preprint submitted to 2024 American Control Conference (ACC)}}
     
    }
    
    \newcounter{thm} 
    \newtheorem{theorem}[thm]{\indent Theorem}
    
    \newtheorem{assumption}{\indent Assumption}
    
    \newtheorem{proposition}{\indent Proposition}
    
    \newtheorem{lemma}{\indent Lemma}
    
    \newtheorem{remark}{\indent Remark}
    
    \newtheorem{corollary}{\indent Corollary}
    
    \newtheorem{definition}{\indent Definition}
    
    \newtheorem{example}{\indent Example}
    
    \newtheorem{fact}{\indent Fact}
    
    \newtheorem{conjecture}{\indent Conjecture}
    
    \newtheorem{experiment}{\indent Experiment}

    \allowdisplaybreaks

    \newlist{enumA}{enumerate}{1}
    \setlist[enumA,1]{label=(A\arabic*),leftmargin=1cm}

    \newlist{enumC}{enumerate}{1}
    \setlist[enumC,1]{label=(C\arabic*),leftmargin=1cm}

    \newcommand{\bd}[0]{\mbox{bd }}

    \usepackage{cite}
    \usepackage{accents}
    
    \newlength\figureheight 
    \newlength\figurewidth

    \allowdisplaybreaks
    
    \graphicspath{ {Figures/} }

    \DeclareMathAlphabet{\mathcal}{OMS}{cmsy}{m}{n} 

    \crefname{equation}{}{}
    
    \usepackage[ruled, linesnumbered, lined]{algorithm2e}
    \SetKwInput{KwInit}{Initialize}
    \SetKwFunction{SafeSet}{SafeSet}
    \SetKwFunction{GetSafe}{GetSafe}
    \SetKwFunction{GetUnsafe}{GetUnsafe}
    
    \SetCommentSty{mycommfont}
    
    \newlist{enumalph}{enumerate}{1}
    \setlist[enumalph]{label=\textit{(\alph*)}}

\begin{document}
\title{Guaranteed-Safe MPPI Through Composite Control Barrier Functions for Efficient Sampling in Multi-Constrained Robotic Systems}
\author{Pedram Rabiee and Jesse B. Hoagg
\thanks{P. Rabiee and J. B. Hoagg are with the Department of Mechanical and Aerospace Engineering, University of Kentucky, Lexington, KY, USA. (e-mail: pedram.rabiee@uky.edu, jesse.hoagg@uky.edu).}
\thanks{This work is supported in part by the National Science Foundation (1849213, 1932105) and Air Force Office of Scientific Research (FA9550-20-1-0028).}
}
\maketitle

\begin{abstract}
We present a new guaranteed-safe model predictive path integral (GS-MPPI) control algorithm that enhances sample efficiency in nonlinear systems with multiple safety constraints. 
The approach use a composite control barrier function (CBF) along with MPPI to ensure all sampled trajectories are provably safe. 
We first construct a single CBF constraint from multiple safety constraints with potentially differing relative degrees, using it to create a safe closed-form control law. 
This safe control is then integrated into the system dynamics, allowing MPPI to optimize over exclusively safe trajectories. 
The method not only improves computational efficiency but also addresses the myopic behavior often associated with CBFs by incorporating long-term performance considerations. 
We demonstrate the algorithm's effectiveness through simulations of a nonholonomic ground robot subject to position and speed constraints, showcasing safety and performance.

\end{abstract}

\section{Introduction}\label{sec:introduction}

The rapid evolution of robotics has led to autonomous systems operating in increasingly complex environments, necessitating control methods capable of handling nonlinear dynamics and multiple constraints.
Model predictive control (MPC) is a powerful tool in this domain \cite{wabersich2022predictive,allgower2012nonlinear,garcia1989model,wabersich2018safe} because it offers a framework for optimizing control inputs over a prediction horizon.
However, traditional MPC approaches often struggle with highly nonlinear systems and with nonlinear constraints; MPC can become computationally prohibitive for real-time applications.
Alternative approaches are the gradient-based methods like the iterative linear quadratic regulator \cite{tassa2012synthesis,li2004iterative}, which are efficient but restrictive in handling multiple nonlinear constraints. 
The challenge is to develop algorithms that balance complex dynamics, multiple constraints, adaptability, and real-time performance without sacrificing safety.

Model-predictive-path-integral (MPPI) control (e.g., \cite{williams2017model,theodorou2010generalized,theodorou2012relative,williams2017information,gandhi2021robust}) is an alternative to traditional MPC.
MPPI  leverages sampling-based techniques and the parallel processing capabilities of modern GPUs. 
Utilizing efficient toolboxes \cite{vlahov2024mppi}, MPPI can simulate a vast number of trajectories simultaneously, enabling it to handle a wide range of cost functions, including non-convex and discontinuous ones, making it particularly suited for complex robotic tasks.
However, MPPI is not without limitations. 
A significant challenge lies in its inability to directly incorporate hard constraints, which are crucial for ensuring safety in robotic applications. 
Additionally, MPPI can suffer from sample inefficiency, particularly when dealing with systems that have strict safety requirements. 
In such cases, a large proportion of sampled trajectories may violate safety constraints, leading to wasted computational effort and potentially compromising the optimization process.

Several approaches have been proposed to address these shortcomings~\cite{kim2022smooth,balci2022constrained,zhang2024multi,yin2022trajectory,yin2023risk,mohamed2022autonomous,williams2018robust}. 
One method involves augmenting the cost function with penalty terms for constraint violations~\cite{yin2023shield}. 
However, this approach can fail to improve sample efficiency significantly because the approach still yields many unsafe trajectories that are explored before converging to safe ones. 
Other techniques attempt to bias the sampling distribution towards safe regions of the state space~\cite{tao2022control}. 
Although these methods can improve performance, they do not provide strict safety guarantees and result in the exploration of unsafe trajectories.

In contrast, CBFs are a computationally efficient tool for safety guarantees \cite{ames2016control,rabiee2023softmax,safari2024time,lindemann2019control,rabiee2023softmin,safari2023time,rabiee2024composition,tan2021high}. 
CBFs are often used in minimum-intervention approaches that minimally modify a nominal control input to maintain safety \cite{ames2016control}. 
While effective for safety, CBFs can exhibit myopic behavior, focusing on immediate safety and sacrificing long-term performance. 
This limitation shifts the burden of optimizing long-term behavior to the nominal control design, which often does not account for safety constraints.
To address this shortcoming,~\cite{yin2023shield} uses a 2-layer design that integrates CBFs with MPPI. 
The approach adds discrete-time CBF constraints to the cost function at each time step of planning, aiming to encourage safety satisfaction throughout the trajectory. 
Then, the approach filters the output using this discrete-time CBF. 
However, this method has several limitations. 
First, it does not significantly improve sample efficiency because the algorithm still explores many unsafe trajectories before converging to safe ones, thus, requiring numerous iterations. 
More importantly, since the safety constraints are not treated as hard constraints, the method may produce trajectories with good performance cost by slightly violating those safety constraints at multiple points. 
Finally, the approach is challenging to tune in the presence of multiple safety constraints.

This paper presents a new guaranteed-safe MPPI (GS-MPPI) control algorithm that improves sample efficiency by considering only guaranteed safe trajectories. 
Our method combines several key ideas to achieve this result. 
First, in \Cref{section:composite CBF}, we utilize the composite CBF technique from~\cite{rabiee2024closed} to construct a single-barrier constraint from multiple safety constraints with differing relative degrees.
This composite CBF constraint is then used in a minimum-intervention architecture to modify a desired control in a minimally invasive manner, resulting in a closed-form safe and instantaneously optimal control.
\Cref{section:mppi} incorporates this closed-form control into the system dynamics to create a guaranteed-safe system. 
Next, we adopt the MPPI formulation from~\cite{williams2017information} to plan for desired control given this  guaranteed-safe system, extending the formulation to accommodate a more general cost function that includes control costs.
Finally,~\Cref{section:mppi cbf} presents the GS-MPPI algorithm.
The key innovation of the approach lies in the use of the guaranteed-safe system in the MPPI framework, and this is enabled by the closed-form control for satisfying multiple safety constraints with potentially differing relative degrees.
Thus, all safety constraints are satisfied as hard constraints at each time step of the planning process. 
This feature results in the generation of only safe trajectories, allowing the MPPI planner to optimize control over guaranteed-safe trajectories exclusively. 
This approach not only improves sample efficiency but also addresses the myopic behavior often associated with CBFs. 
Specifically, using MPPI to design the desired control and accounting for the existence of inner-loop minimum-intervention CBF control, the approach enables long-term performance optimization with safety guarantees.
We demonstrate this new method in simulations of a nonholonomic ground robot subject to position and speed constraints.

\section{Notation}
The interior, boundary, and closure of $\SA \subseteq \BBR^n$ are denoted by $\mbox{int }\SA$, $\bd\SA$, $\mbox{cl }\SA$, respectively.

Let $\eta:\BBR^n \to \BBR^\ell$ be continuously differentiable. 
Then, $\eta^\prime :\BBR^n \to \BBR^{\ell \times n}$ is defined as $\eta^\prime(x) \triangleq \pderiv{\eta(x)}{x}$. 
The Lie derivatives of $\eta$ along the vector fields of $\psi:\BBR^n \to \BBR^{n \times m}$ is defined as
\begin{equation*}
L_\psi \eta(x) \triangleq \eta^\prime(x) \psi(x).
\end{equation*}
If $m=1$, then for all positive integers $d$, define
\begin{equation*}
    L_\psi^d \eta(x) \triangleq L_\psi L_\psi^{d-1} \eta(x).
\end{equation*} 
Throughout this paper, we assume that all functions are sufficiently smooth such that all derivatives that we write exist and are continuous.

A continuous function $a \colon \BBR \to \BBR$ is an \textit{extended class-$\SK$ function} if it is strictly increasing and $a(0)=0$.

Let $\rho>0$, and consider $\mbox{softmin}_\rho : \BBR^N\to \BBR$ defined by 
\begin{equation}\label{eq:softmin}
\mbox{softmin}_\rho (z_1,\ldots,z_N) \triangleq -\frac{1}{\rho}\log\sum_{i=1}^Ne^{-\rho z_i},
\end{equation}
which is the log-sum-exponential \textit{soft minimum}.

Let $\Omega$ be a sample space and $\BBP$ be a probability measure defined on $\Omega$. Let $X : \Omega \to \mathbb{R}$ be a random variable with probability density function $p$ with respect to the measure $\BBP$. The expectation of $X$, assuming it exists, is defined as
\begin{equation}
\mathbb{E}_\BBP[X] = \int_{\mathbb{R}} x p(x) dx.
\end{equation}

Let $\Omega$ be a sample space and let $\BBP$ and $\BBQ$ be two probability measures defined on $\Omega$. 
Suppose $p$ and $q$ are the probability density function of $\BBP$ and $\BBQ$, respectively. The Kullback-Leibler (KL) divergence from $\BBQ$ to $\BBP$ is defined as
\begin{equation}
\BBD_{\text{KL}}(\BBP \| \BBQ) = \int_{X} p(x) \log\left(\frac{p(x)}{q(x)}\right) dx,
\end{equation}
assuming that $p$ is absolutely continuous with respect to $q$ and the integral is well-defined.

\section{Problem Formulation} \label{section:problem_formulation}
Consider 
\begin{equation}\label{eq:dynamics}
\dot x(t) = f(x(t))+g(x(t)) u(t),
\end{equation}
where $x(t) \in \BBR^{n}$ is the state, $x(0) = x_0 \in \BBR^n$ is the initial condition, $f:\BBR^n \to \BBR^n$ and $g:\BBR^n\to \BBR^{n\times m}$ are locally Lipschitz continuous on $\BBR^n$, and $u: [0,\infty) \to \BBR^m$ is the control, which is locally Lipschitz continuous on $\BBR^n$.

Let  $h_1,h_2,\ldots,h_\ell \colon \BBR^n \to \BBR$ be continuously differentiable, and for all $\jmath \in \{ 1,2,\ldots,\ell\}$, define 
\begin{equation}
    \SC_{\jmath,0} \triangleq \{ x \in \BBR^n \colon h_\jmath(x)\geq 0 \}. 
\end{equation}
The \textit{safe set} is
\begin{equation}\label{eq:S_s}
\SSS_\rms \triangleq {\bigcap_{\jmath=1}^\ell} \SC_{\jmath,0}.
\end{equation}
Unless otherwise stated, all statements in this paper that involve the subscript $\jmath$ are for all $\jmath \in \{ 1,2,\ldots,\ell\}$.
We make the following assumption:
\begin{enumA}

\item There exists a positive integer $d_\jmath$ such that for all $x \in \BBR^n$ and all $i \in \{ 0,1,\ldots,d_\jmath-2\}$, $L_g L_f^ih_\jmath(x) = 0$; and for all $x \in \SSS_\rms$, $L_g L_f^{d_\jmath-1}h_\jmath(x) \neq 0$. \label{cond:hocbf.a}
\end{enumA}

Assumption~\ref{cond:hocbf.a} implies $h_\jmath$ has well-defined relative degree $d_\jmath$ with respect to~\eqref{eq:dynamics} on $\SSS_\rms$; however, relative degrees $d_1,\ldots,d_\ell$ need not be equal.

Let $T > 0$ be the time horizon, $\hat \phi: \BBR^n  \to \BBR$ be the terminal cost function,
$\hat \psi: \BBR^n \times \BBR^m \to \BBR$ be the running cost function and define the cost functional 

\begin{align}\label{eq:cost}
\hat \SJ(u) &\triangleq \hat \phi(x(T))+ \int_0^{T}\hat \psi(x(t), u(t)) \dt.
\end{align}

The objective is to find a control law $u: [0,T] \to \BBR^m$ that minimizes the cost functional $\hat \SJ(u)$ subject to~\eqref{eq:dynamics} and the safety constraint that for all $t\in[0,T]$, $x(t)\in \SSS_\rms$.

\section{Guaranteed-Safe Model Predictive Path Integral}
\begin{figure}[t!]
\center{\includegraphics[width=0.4\textwidth,clip=true,trim= 0.0in 0.0in 0in 0.0in] {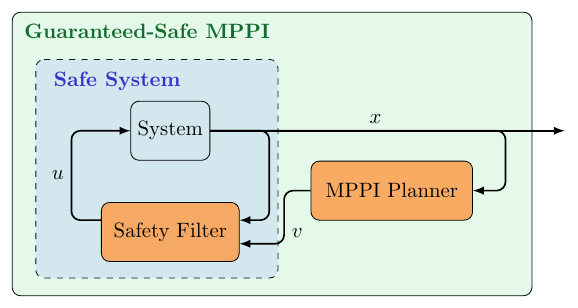}}
\caption{Block diagram of the Guaranteed-Safe Model Predictive Path Integral (MPPI) control architecture, illustrating the interaction between the Safety Filter, Safe System, and MPPI Planner components.} 
\label{fig:block_diag1}
\end{figure}

\Cref{fig:block_diag1} illustrates the architecture of the proposed method for guaranteed-safe MPPI.
The method consists of 3 main components: 
\begin{enumerate}
    \item \textit{Safety Filter:} In \Cref{section:composite CBF}, we first design a closed-form safe and instantaneously optimal control from the safety constraints. This procedure follows the method in~\cite{rabiee2024closed}. Given a \textit{desired control} $v$, the safety filter provides the minimally invasive control—a control that is the closest to $v$ while satisfying the safety constraints. Thus, it is both safe and optimal in the sense that it is as close as possible to a desired control.
    
    \item \textit{Safe System:} In Section \ref{section:mppi}, we incorporate this closed-form control into the dynamics \eqref{eq:dynamics}, creating a safe system that accepts arbitrary desired control while ensuring:
    \textit{(a)} The underlying system's safety is maintained, i.e., $x(t) \in \SSS_\rms$ for all $t \ge 0$,
    \textit{(b)} The implemented control is as close as possible to the desired control $v$.

    \item \textit{MPPI Planner:} Also described in Section \ref{section:mppi}, we use this safe system to design the desired control $v$ using an MPPI planner. As the system is inherently safe, all trajectories generated by the MPPI planner are guaranteed to be safe.
\end{enumerate}    

\subsection{Closed-Form Safe and Instantaneously Optimal Control}
\label{section:composite CBF} 

In this section, we review and adapt the method presented in \cite{rabiee2024closed} for constructing a single composite Relaxed Control Barrier Function (R-CBF) from multiple CBFs with potentially different relative degrees. We then utilize this composite soft-minimum R-CBF to derive a closed-form optimal control law that guarantees safety while minimally deviating from a desired control input.
The following definition is needed.

\begin{definition}\label{def:RCBF}
\rm
Let $\eta \colon \BBR^n \to \BBR$ be continuously differentiable, and define $\SD \triangleq \{ x \in \BBR^n \colon \eta(x) \ge 0 \}$.
Then, $\eta$ is a \textit{relaxed control barrier function} (R-CBF) for~\cref{eq:dynamics} on $\SD$ if for all $x \in \mbox{bd }\SD$,
    \begin{equation}
        \sup_{u\in \BBR^{m}} L_f \eta(x) + L_g \eta(x)u \ge 0.\label{def:RCBF.1}
    \end{equation} 
\end{definition}
The following result provides sufficient conditions for $\eta$ to be an R-CBF. The result also provides sufficient conditions for the zero-superlevel set of $\eta$ to be control forward invariant.
\begin{lemma}\label{lem:rcbf}
\rm 
\cite[Lemma 1]{rabiee2024closed}
Let $\eta \colon \BBR^n \to \BBR$ be continuously differentiable, and define $\SD \triangleq \{ x \in \BBR^n \colon \eta(x) \ge 0 \}$. 
Assume that for all $x \in \mbox{bd }\SD$, if $L_f \eta(x) \le 0$, then $L_g \eta(x) \ne 0$. 
Then, the following statements hold:
\begin{enumalph}
\item \label{lem:rcbf.1}
$\eta$ is an R-CBF for~\eqref{eq:dynamics} on $\SD$. The following result provides sufficient conditions for $\eta$ to be an R-CBF.
The result also provides sufficient conditions for the zero-superlevel set of $\eta$ to be control forward invariant. 
\item \label{lem:rcbf.2}
If $\eta^\prime$ is locally Lipschitz on $\SD$, then $\SD$ is control forward invariant with respect to~\eqref{eq:dynamics}.
\end{enumalph}
\end{lemma}

Let $b_{\jmath,0}(x) \triangleq h_\jmath(x)$. 
For $i\in\{0, 1\ldots,d_\jmath-2\}$, let $\alpha_{\jmath,i} \colon \BBR \to \BBR$ be a locally Lipschitz extended class-$\SK$ function, and consider $b_{\jmath, i+1}:\BBR^n \to \BBR$ defined by
\begin{equation}\label{eq:b_def}
    b_{\jmath,i+1}(x) \triangleq L_f b_{\jmath,i}(x) + \alpha_{\jmath,i}(b_{\jmath,i}(x)).
\end{equation}
For $i \in \{1\ldots,d_\jmath-1\}$, define
\begin{equation}\label{eq:C_j_i_def}
    \SC_{\jmath,i} \triangleq \{x\in \BBR^n: b_{\jmath,i}(x) \ge 0\}.
\end{equation}
Next, define
\begin{gather}  \label{eq:C_j_def}
    \SC_\jmath \triangleq 
        \begin{cases}
        \SC_{\jmath,0}, & d_\jmath = 1,\\
    \bigcap_{i=0}^{d_\jmath - 2} \SC_{\jmath,i}, & d_\jmath > 1,
    \end{cases}
\end{gather}
and 
\begin{equation}  
    \SC \triangleq \bigcap_{\jmath=1}^{\ell} \SC_\jmath. \label{eq:C_def}
\end{equation}
Note that $\SC \subseteq \SSS_\rms$.
In addition, note that if $d_1,\ldots,d_\ell \in \{1,2\}$, then $\SC = \SSS_\rms$.

Let $\rho > 0$, and consider the candidate R-CBF $h:\BBR^n \to \BBR$ defined by
\begin{equation}\label{eq:h_def}
    h(x) \triangleq \mbox{softmin}_\rho \Big ( b_{1, d_1-1}(x), b_{2, d_2-1}(x), \ldots,b_{\ell, d_\ell-1}(x) \Big ). 
\end{equation}
The zero-superlevel set of $h$ is
\begin{equation}\label{eq:defS_softmin}
\SH \triangleq \{ x \in \BBR^n \colon h(x) \ge 0 \}.
\end{equation}
Proposition 2 in~\cite{rabiee2024closed} demonstrates that $\SH \subseteq \bigcap_{\jmath=1}^{\ell} \SC_{\jmath,d_\jmath-1}$.
Note that $\SH$ is not generally a subset of $\SSS_\rms$ or $\SC$. 
In the special case where $d_1=\cdots=d_\ell = 1$, it follows that $\SH \subseteq \SC = \SSS_\rms$, and as $\rho \to \infty$, $\SH \to \SSS_\rms$. 
Next, we define 
\begin{equation}\label{eq:S}
    \SSS \triangleq \SH \cap \SC,
\end{equation}
and since $\SC \subseteq \SSS_\rms$, it follows that $\SSS \subseteq \SSS_\rms$.
Next, define
\begin{align}
    \SB &\triangleq \{ x \in \bd \SH \colon L_fh(x) \le 0 \}\nn\\
    &=\{ x \in \BBR^n \colon L_fh(x) \le 0 \mbox{ and } h(x) =0 \}. \label{eq:SB}
\end{align}
\Cref{lem:rcbf} implies that if $L_gh$ is nonzero on $\SB$, then $h$ is an R-CBF.
\Cref{lem:rcbf} also implies that if $L_gh$ is nonzero on $\SB$ and $h^\prime$ is locally Lipschitz, then $\SH$ is control forward invariant.
Proposition 3 in~\cite{rabiee2024closed} shows that in this case not only is $\SH$ control forward invariant but $\SSS$ is also control forward invariant.

Next, we use the composite soft-minimum R-CBF \eqref{eq:h_def} to construct a closed-form optimal control that guarantees safety.
Specifically, we design an optimal control subject to the constraint that $x(t) \in \SSS \subseteq \SSS_\rms$.

Let $\alpha:\BBR \to \BBR$ be a locally Lipschitz nondecreasing function such that $\alpha(0) = 0$, and consider the R-CBF safety constraint function $\omega:\BBR^n\times \BBR^m \times \BBR \to \BBR$ given by
\begin{equation}\label{eq:omega_def}
\omega(x, \hat u, \hat \mu) \triangleq L_f h(x) + L_g h(x) \hat u + \alpha (h(x)) + \hat \mu h(x)
\end{equation}
where $\hat u$ is the control variable; and $\hat \mu$ is a slack variable.
Next, let $v : \BBR^n \to \BBR^m$ denote a \textit{desired control} designed to satisfy performance specifications, which can be independent of and potentially conflict with safety. Let $\gamma > 0$, and consider the minimum-intervention cost function $\SJ_{\rm MI}:\BBR^n\times \BBR^m \times \BBR \times \BBR^m \to \BBR$ given by
\begin{equation}
    \SJ_{\rm MI}(x, \hat u, \hat \mu, v) \triangleq \frac{1}{2}\|\hat  u - v \|^2 + \frac{\gamma}{2} \hat \mu^2.
\end{equation}
The objective is to find the $(\hat u,  \hat \mu)$ that minimizes the cost $\SJ_{\rm MI}(x, \hat u, \hat \mu)$ subject to the R-CBF safety constraint $\omega(x, \hat u, \hat \mu) \ge 0$.

Proposition 4 in~\cite{rabiee2024closed} shows that if for all $x \in \SB$, $L_gh(x) \ne 0$, then this problem is feasible on $\BBR^n$.
Then, for all $x \in \BBR^n$, the solution $u_*:\BBR^n\to \BBR^m$ is given by
\begin{align}\label{eq:safe_control}
u_*(x, v) &= v +  \frac{L_gh(x)^\top \max\{0,-\omega(x, v, 0)\}}{L_gh(x)L_gh(x)^\top + \gamma^{-1} h(x)^2}.
\end{align}
The next theorem is the main result on safety using this control.

\begin{theorem}\label{thm:smocbf}\rm
\cite[Theorem 2]{rabiee2024closed}
Consider~\eqref{eq:dynamics}, where \ref{cond:hocbf.a} is satisfied, and consider $u$ given by~\Cref{eq:b_def,eq:h_def,eq:safe_control}, with $v$ locally Lipschitz continuous on $\BBR^n$.
Assume that $h^\prime$ is locally Lipschitz on $\SH$, and for all $x \in \SB$, $L_gh(x) \ne 0$.
Let $x_0\in \SSS$.
Then, for all $t \in I(x_0)$, $x(t) \in \SSS \subseteq \SSS_\rms$.
\end{theorem}

Theorem \ref{thm:smocbf} establishes the safety guarantees for the control law $u$ given by~\Cref{eq:b_def,eq:h_def,eq:safe_control}. This result provides a control law that ensures the system remains in the safe set $\SSS_\rms$ while minimally deviating from the desired control $v$.
We can now proceed to design an optimal control strategy using MPPI methods, which we will explore in the next subsection.

\subsection{Safe Model Predictive Path Integral}
\label{section:mppi}
In this section, we mirror the approach outlined in \cite{williams2017information} to develop our safe MPPI method. However, we extend their formulation by modifying the cost function to allow for control-dependent costs.

To begin, we construct a safe dynamical system by incorporating the control law from \Cref{eq:b_def,eq:h_def,eq:safe_control} into the original dynamics \cref{eq:dynamics}. This integration yields a new system that inherently satisfies the safety requirements, allowing us to solve the optimization problem while implicitly respecting safety constraints. The resulting safe dynamics are given by
\begin{equation}\label{eq:safe_dyn}
    \dot x(t) = f(x(t))+g(x(t))u_*(x(t), v(t)).
\end{equation}

To facilitate the MPPI derivation, we discretize \eqref{eq:safe_dyn} using the Euler approximation method.
Let $k \in \{0, 1, ..., N-1\}$ denote the discrete time index, where $N$ is the number of time steps in the planning horizon. The discrete-time safe dynamics are given by
\begin{equation}\label{eq:dynamics_discrete}
x_{k+1} = F(x_k, v_k),
\end{equation}
where $x_k\in\BBR^n$, $v_k\in\BBR^m$, and $F:\BBR^n\times\BBR^m \to \BBR^n$ is defined as
\begin{align}\label{eq:F_def}
F(x_k, v_k) \triangleq x_k + [f(x_k) + g(x_k)u_*(x_k, v_k)]T_\rms
\end{align}
where $T_\rms>0$ represents the time step between discrete samples.

We model the desired control as a sequence of random variables $\{v_k\}_{k=0}^{N-1}$, where each $v_k$ is sampled from a normal distribution
\begin{equation}\label{eq:u_d}
v_k \sim \SN(\mu_k, \Sigma), \quad k\in \{0, 1,\ldots, N-1\}.
\end{equation}
where $\mu_k \in \BBR^m$ represent the mean control input at time step $k$, and $\Sigma\in\BBR^{m\times m}$ is a positive definite matrix covariance matrix. This stochastic formulation captures inherent system uncertainties and facilitates state space exploration.

Let $M = (\mu_0, \mu_1, ..., \mu_{N-1}) \in \BBR^{m \times N}$ denote the mean control sequence, and $V = (v_0, v_1, ..., v_{N-1}) \in \BBR^{m \times N}$ represent the sequence of control inputs. The choice of $M$ determines the probability distribution of $V$.

Let $\BBP_M$ denote the probability measure of $V$ given $M$. The corresponding density function $p_M: \BBR^{m \times N} \to [0,1]$ is defined as
\begin{equation}\label{eq:traj_dist}
    p_M(V) = Z \exp\left(-\frac{1}{2} \sum_{k=0}^{N-1} (v_k - \mu_k)^\top\Sigma^{-1}(v_k-\mu_k)\right)
\end{equation}
where $Z=((2\pi)^m |\Sigma|)^{-\frac{N}{2}}$ is the normalizing constant.

Let $x_0\in \SSS_\rms$, be the initial state. Define the discrete-time cost function $\SJ: \BBR^{m \times N} \to \BBR$ as
\begin{equation}\label{eq:cost_discrete}
    \SJ(V) \triangleq \phi(x_N) +\sum_{k=1} ^{N-1}\psi(x_k, v_k),
\end{equation}
where $(x_1, \ldots, x_N)$ are related to $V$ and $x_0$ by \cref{eq:dynamics_discrete,eq:F_def}, $\phi: \BBR^n \to \BBR$ is the terminal cost function, and $\psi: \BBR^n \times \BBR^m \to \BBR$ is the running cost function.

We formulate our control objective as finding a probability measure $\BBQ$ over control sequences that minimizes
\begin{equation}\label{eq:mppi_obj}
\BBE_{\BBQ}[\SJ(V)] + \lambda \BBD_{\rm KL}(\BBQ\|\BBP_0)
\end{equation}
where $\BBP_0$ is the probability measure of the uncontrolled system (i.e., $M\equiv 0$) with density function $p_0$, and $\lambda > 0$ is a temperature parameter.
By minimizing this objective, we seek a control distribution that not only reduces the expected cost but also maintains some similarity to the uncontrolled system dynamics. This approach encourages exploration in the control space, while simultaneously preventing the optimal distribution from deviating too far from the natural system behavior. The parameter $\lambda$ controls the trade-off between 
minimizing the expected cost $\BBE_{\BBQ}[\SJ(V)]$ (exploitation) and minimizing the deviation from the uncontrolled dynamics as measured by $\BBD_{\rm KL}(\BBQ\|\BBP_0)$ (exploration).

Following the approach in \cite{williams2017information}, the optimal probability measure $\BBP_*$ that minimizes \eqref{eq:mppi_obj} has a density function $p_*(V)$ given by
\begin{equation}\label{eq:optimal_dist}
p_*(V) = \frac{\exp\left(-\frac{1}{\lambda} \SJ(V)\right)}{\BBE_{\BBP_0}\left[\exp\left(-\frac{1}{\lambda}\SJ(V)\right)\right]}p_0(V).
\end{equation}

It is important to note that the optimal probability measure $\BBP_*$ does not necessarily have the same structure as our parameterized family of distributions $\BBP_M$. Each $\BBP_M$ in this family is constrained to be a multivariate normal distribution with mean sequence $M$ and fixed covariance matrix $\Sigma$. Our goal is to find the optimal mean control sequence $M_*\in \BBR^{m\times N}$ such that $\BBP_{M_*}$ is as close as possible to $\BBP_*$ within this parameterized family.
To achieve this, we determine the optimal mean sequence $M_*$ by minimizing the KL divergence between $\BBP_*$ and $\BBP_M$. Formally, we express this optimization problem as
\begin{equation}\label{eq:KL}
M_* \triangleq \underset{M\in\BBR^{m\times N}}{\mbox{argmin}} \, \BBD_{\rm KL}(\BBP_* \| \BBP_M).
\end{equation}

This formulation allows us to find the best approximation of $\BBP_*$ within the constrained structure of our parameterized distribution $\BBP_M$.

To derive the update law for $M_*$, we begin by expanding the KL divergence expression from \cref{eq:KL} as
\begin{align}
    &\BBD_{\rm KL}(\BBP_* \| \BBP_M) = \int \log\left(\frac{p_*(V)}{p_M(V)} \right)p_*(V)\, \rmd V\nn\\
    &\quad=\int \log\left(\frac{p_*(V)}{p_0(V)}\frac{p_0(V)}{p_M(V)} \right)p_*(V)\, \rmd V\nn\\
    &\quad=\int \log\left(\frac{p_*(V)}{p_0(V)} \right)p_*(V)\, \rmd V\nn\\
    &\quad\qquad- \int \log\left(\frac{p_0(V)}{p_M(V)} \right)p_*(V)\, \rmd V.\label{eq:temp1}
\end{align}
Since the first integral in~\cref{eq:temp1} is independent of $M$, we can reformulate the optimization problem in \cref{eq:KL} as
\begin{equation}\label{eq:mu_optim_new}
M_* = \underset{M\in\BBR^{m\times N}}{\mbox{argmax}} \, \int \log\left(\frac{p_0(V)}{p_M(V)} \right)p_*(V)\, \rmd V.
\end{equation}
Note from~\cref{eq:traj_dist} that
\begin{align}\label{eq:dist_ratio}
  \frac{p_M(V)}{p_0(V)} =
  \exp\Bigg(&\sum_{k=0}^{N-1} \mu_k^\top \Sigma^{-1} v_k -\frac{1}{2} \mu_k^\top \Sigma^{-1} \mu_k\Bigg),
\end{align}
where $p_0$ is obtained by setting $M= (0_m,\ldots, 0_m)$ in~\cref{eq:traj_dist}.
Substituting this in~\cref{eq:mu_optim_new} yields
\begin{equation}\label{eq:temp4}
M_* = \underset{M\in\BBR^{m\times N}}{\mbox{argmax}} \int \Bigg(\sum_{k=0}^{N-1} \mu_k^\top \Sigma^{-1} v_k -\frac{1}{2} \mu_k^\top \Sigma^{-1} \mu_k\Bigg) p_*(V) \rmd V.
\end{equation}
Since $M$ does not depend on $\BBP_*$, we can separate the terms in \cref{eq:temp4} and use the fact that $\int p_*(V) \, \rmd V= 1$. This allows us to rewrite~\cref{eq:mu_optim_new} as
\begin{align}\label{eq:KL_simplified}
M_* = \underset{M\in\BBR^{m\times N}}{\mbox{argmax}} \,&-\frac{1}{2} \sum_{k=0}^{N-1}\mu_k^\top \Sigma^{-1} \mu_k\nn\\
&+\sum_{k=0}^{N-1} \mu_k^\top \Bigg(\int \Sigma^{-1} v_k p_*(V) \, \rmd V \Bigg).
\end{align}
The expression in \cref{eq:KL_simplified} is concave with respect to each $\mu_k$. To find its maximum, we take the gradient with respect to each $\mu_k$ and set it to zero. Solving for $\mu_k$, we obtain the optimal mean control
\begin{equation}\label{eq:mu_star_final}
    \mu_{*k}= \int v_k \, p_*(V)\, \rmd V = \BBE_{\BBP_*}[v_k].
\end{equation}
The result in \cref{eq:mu_star_final} provides us with the optimal mean control $\mu_{*k}$ as the expected value of $v_k$ under the optimal distribution $\BBP_*$.
However, we cannot directly sample from $\BBP_*$, so we compute this integral using importance sampling with a proposal distribution $\BBP_M$ as followed
\begin{align}
    \mu_{*k} &=\int  v_k \frac{p_*(V)}{p_0(V)}\frac{p_0(V)}{p_M(V)} p_M(V) \,\rmd V \nn\\
    & = \BBE_{\BBP_M}\left[v_k \frac{p_*(V)}{p_0(V)}\frac{p_0(V)}{p_M(V)} \right]\label{eq:mu_star_exp}.
\end{align}
From~\cref{eq:optimal_dist,eq:dist_ratio}, the importance sampling weight can be written as
\begin{align}
    \frac{p_*(V)}{p_0(V)}\frac{p_0(V)}{p_M(V)} &= 
    \frac{1}{\eta} \exp\Bigg(-\frac{1}{\lambda}\SJ(V) + \sum_{k=0}^{N-1} \mu_k^\top \Sigma^{-1} v_k \nn\\
    &\qquad\qquad\quad-\frac{1}{2} \mu_k^\top \Sigma^{-1} \mu_k    
    \Bigg),\label{eq:temp3}
\end{align}
where $\eta \triangleq \BBE_{\BBP_0}\left[\exp\left(-\frac{1}{\lambda}\SJ(V)\right)\right]$. 
For each $k \in \{0, 1, \ldots, N-1\}$, let $\varepsilon_k \sim \mathcal{N}(0, \Sigma)$ be independent Gaussian random variables, and define $v_k = \mu_k + \varepsilon_k$. We denote the sequence of these random variables as $\SE = (\varepsilon_0, \varepsilon_1, \ldots, \varepsilon_{N-1}) \in \BBR^{m \times N}$.
Using this change of variables, we can rewrite the importance sampling weight from \cref{eq:temp3} as
\begin{equation}\label{eq:temp6}
\frac{p_*(V)}{p_0(V)}\frac{p_0(V)}{p_M(V)} =
\frac{1}{\eta} \exp\left(-\frac{1}{\lambda} S(M, \SE)\right),
\end{equation}

where $S: \BBR^{m \times N} \times \BBR^{m \times N} \to \BBR$ is defined as
\begin{align}\label{eq:S_cost}
S(M, \SE) &\triangleq \SJ(M+\SE) + \frac{\lambda}{2} \sum_{k=0}^{N-1} \mu_k^\top\Sigma^{-1}(\mu_k + 2\varepsilon_k).
\end{align}

Moreoever, we can express $\eta$ as
\begin{align}
 \eta &= \BBE_{\BBP_M}\left[\exp\left(-\frac{1}{\lambda}\SJ(V)\right) \frac{p_0(V)}{p_M(V)}\right]\nn\\
 &=\BBE_{\BBP_M}\left[\exp\left(-\frac{1}{\lambda} S(M, \SE)\right)\right]\label{eq:temp5}.
\end{align}

Substituting~\cref{eq:S_cost,eq:temp5,eq:temp6} into~\eqref{eq:mu_star_exp} we obtain
\begin{align*}
\mu_{*k} = \frac{\BBE_{\BBP_M}\left[(\mu_k+\varepsilon_k) \exp\left(-\frac{1}{\lambda} S(M, \SE)\right) \right]}{\BBE_{\BBP_M}\left[\exp\left(-\frac{1}{\lambda} S(M, \SE)\right) \right]}.
\end{align*}
Separating the terms in the numerator inside the expectation yields
\begin{align}\label{eq:mu_*_update_law}
\mu_{*k} = \mu_k + \frac{\BBE_{\BBP_M}\left[ \exp\left(-\frac{1}{\lambda} S(M, \SE)\right) \varepsilon_k\right]}{\BBE_{\BBP_M}\left[\exp\left(-\frac{1}{\lambda} S(M, \SE)\right) \right]}.
\end{align}

The expression \cref{eq:mu_*_update_law} provides an update rule for the optimal mean control $\mu_{*k}$. It shows that the optimal mean is the current mean plus a correction term. This theoretical framework provides the foundation for our guaranteed-safe MPPI approach. 

The MPPI approach addresses the original control problem by discretizing the dynamics and reformulating the cost. While the discrete-time cost $\SJ(V)$ approximates the continuous-time cost $\hat \SJ(u)$, our MPPI formulation in \eqref{eq:mppi_obj} introduces an additional term $\lambda \BBD_{\rm KL}(\BBQ|\BBP_0)$. This term, reflected in the final cost $S(M, \SE)$ in \eqref{eq:S_cost}, balances exploration and exploitation, with $\lambda$ controlling this trade-off. This formulation allows us to solve the original problem while incorporating the benefits of stochastic optimization, all within a framework that inherently respects the safety constraints defined by $\SSS_\rms$.
Next section develops a practical algorithm based on these results.

\section{MPPI-CBF: An Algorithm for Guaranteed-Safe Model Predictive Path Integral}\label{section:mppi cbf}

In the preceding section, we have established a framework for guaranteed-safe MPPI. We first developed a composite R-CBF that ensures safety for multiple constraints, potentially with different relative degrees. This allowed us to construct a safe dynamical system that inherently respects these constraints. We then adapted the MPPI method to this safe system, deriving an optimal control law that balances performance optimization with safety guarantees. However, to implement this approach in practical settings, we need to translate these theoretical insights into a concrete algorithm. In the following section, we present MPPI-CBF, an algorithm that realizes our guaranteed-safe MPPI framework.

To implement our guaranteed-safe MPPI framework in practical settings, we address two key aspects: the approximation of the expectation in the update law~\cref{eq:mu_*_update_law}, and the receding horizon implementation. These are realized in Algorithms \ref{alg:mppi} and \ref{alg:mppi_cbf}.

First, to make the computation tractable, we approximate the expectation in \cref{eq:mu_*_update_law} using a finite number of trajectories $K$. This leads to the following approximated update law

\begin{align}\label{eq:mu_star_sampled}
\mu_{*k} \approx \mu_k + \frac{\sum_{j=1}^K \exp\left(-\frac{1}{\lambda} S(M, \SE^{(j)})\right) \varepsilon_k^{(j)}}{\sum_{j=1}^K\exp\left(-\frac{1}{\lambda} S(M, \SE^{(j)})\right) },
\end{align}
where the subscript $(j)$ denotes the $j$-th sampled trajectory.

Second, we employ a receding horizon approach. While the original problem has a horizon $T$, for large $T$, we use a shorter horizon $N T_\rms < T$, where $N$ is the number of time steps and $T_\rms$ is the planning time step.

Algorithm \ref{alg:mppi} describes the MPPI Planner. Given the initial state $x_0$ and a mean control sequence $M$, it computes the optimal mean control sequence using the MPPI update law \cref{eq:mu_star_sampled} for the current horizon. The algorithm returns the optimal mean control sequence and the first desired input corresponding to the best trajectory explored by the MPPI planner. We use this best trajectory (taking a greedy approach) instead of sampling from the optimized distribution. This choice ensures that, even after filtering the desired controls, the resulting safe trajectory will still have a low cost, as it's derived from the input that yielded the lowest cost during exploration.

Algorithm \ref{alg:mppi_cbf} implements the overall GS-MPPI framework, featuring a two-level control structure.
The outer loop, operating at a coarser time step $T_\rms$, calls the MPPI Planner to generate desired control sequences. The inner loop, using a finer time step $\delta t$, applies the safety filter \cref{eq:safe_control} to the desired control. This dual-timescale approach allows for faster, more frequent safety filtering while maintaining a computationally feasible planning horizon.
At each planning step, the algorithm calls the MPPI Planner, executes the filtered control, updates the state, and shifts the control sequence for the next iteration. This process repeats until the full time horizon $T$ is covered, ensuring continuous safety while optimizing the trajectory over an extended horizon.

\begin{algorithm}[ht]\label{alg:mppi}
\DontPrintSemicolon
\caption{MPPI Planner}
\SetKwData{NumSamples}{$K$}
\SetKwData{Lambda}{$\lambda$}
\SetKwData{Nhorizon}{$N$}
\SetKwData{covariance}{$\Sigma$}
\SetKwComment{Comment}{$\triangleright$\ }{}
\Comment{Global:\Nhorizon, \NumSamples, \Lambda, \covariance, $T_\rms$}
\KwIn{$x_0$: initial state, $M$: mean control sequence}
\KwOut{$M_*$: optimal mean control sequence, $v_0^{(j_{\rm best})}$: first control of the lowest cost trajectory}\SetKwFunction{FMPPI}{MPPI\_Planner}
\SetKwProg{Fn}{Function}{:}{}
\Fn{\FMPPI{$x_0$, $M$}}{
    \Comment{\ProgSty{Generate $K$ trajectories and compute their costs}}
    
    \For{$j \leftarrow 1$ \KwTo $K$}{
        $\SJ^{(j)} \leftarrow 0$\;
        $x \leftarrow x_0$\;
        \For{$k \leftarrow 0$ \KwTo $N-1$}{
            $\varepsilon^{(j)}_k \sim \mathcal{N}(0, \Sigma)$\;
            $v_k^{(j)} \leftarrow \mu_k + \varepsilon^{(j)}_k$\;
            $\SJ^{(j)} \leftarrow \SJ^{(j)} + \psi(x, v_k^{(j)})$\;
            $x \leftarrow F(x, v_k^{(j)})$\;
        }
        $\SJ^{(j)} \leftarrow \SJ^{(j)} + \phi(x)$\;
    }
    \Comment{\ProgSty{Compute importance sampling weights}}
    $\xi \leftarrow \min\{\SJ^{(1)}, \SJ^{(2)}, \ldots, \SJ^{(K)}\}$\;
    $j_{\rm best} \leftarrow \rm{argmin}\{\SJ^{(1)}, \SJ^{(2)}, \ldots, \SJ^{(K)}\}$\;
    $\eta \leftarrow \sum_{j=1}^{K} \exp(-\frac{1}{\lambda}(\SJ^{(j)} - \xi))$\;
    \For{$j \leftarrow 1$ \KwTo $K$}{
        $w^{(j)} \leftarrow \frac{1}{\eta} \exp(-\frac{1}{\lambda}(\SJ^{(j)} - \xi))$\;
    }
    \Comment{\ProgSty{Update mean control sequence}}
    \For{$k \leftarrow 0$ \KwTo $N-1$}{
        $\mu_k \leftarrow \mu_k + \sum_{j=1}^{K} w^{(j)} \varepsilon_k^{(j)}$\;
    }
    $M_*\leftarrow(\mu_0, \mu_1, \ldots, \mu_{N-1})$\;
\Return{$M_*$, $v_0^{(j_{\rm best})}$}
}
\textbf{End Function}
\end{algorithm}

\begin{algorithm}[h!]\label{alg:mppi_cbf}
\DontPrintSemicolon
\caption{GS-MPPI: An Algorithm for Guaranteed-Safe Model Predictive Path Integral}
\SetKwData{FDyn}{$F$}
\SetKwData{GDyn}{$G$}
\SetKwData{USafe}{$u_*$}
\SetKwComment{Comment}{$\triangleright$\ }{}
\Comment{Global: $T_\rms$, $\delta t$, $N$}
\KwIn{$x_0$: initial state, $T$: total simulation time}
\KwOut{State and control trajectories}

Initialize $M \leftarrow (\mu_0, \mu_1, \ldots, \mu_{N-1})$\;
$x \leftarrow x_0$\;
$n_{\rm inner} \leftarrow \lfloor T_\rms / \delta t \rfloor$\;

\For{$t = 0$ \KwTo $T$ \textbf{with step} $T_\rms$}{
    \Comment{\ProgSty{Plan optimal desired control sequence}}
    $M, v \leftarrow \FuncSty{\text{MPPI\_Planner}}(x, M)$\;
    \Comment{\ProgSty{Execute desired control and update state}}
    \For{$i = 1$ \KwTo $n_{\rm inner}$}{
        $u_{\rm safe} \leftarrow \USafe(x, v)$ \Comment*[r]{Eq. \cref{eq:safe_control}}
        $x \leftarrow$ Evolution of \eqref{eq:dynamics} under $u_{\rm safe}$ after $\delta t$ \;        
    }
    \Comment{\ProgSty{Shift mean control sequence for next step}}
    \For{$k = 0$ \KwTo $N-2$}{
        $\mu_k \leftarrow \mu_{k+1}$\;
    }
    $\mu_{N-1} \leftarrow \text{Initialize}()$
}
\end{algorithm}
 
\section{Numerical Examples}
\begin{figure}[t!]
\center{\includegraphics[width=0.49\textwidth,clip=true,trim= 0.in 0.in 0in 0in] {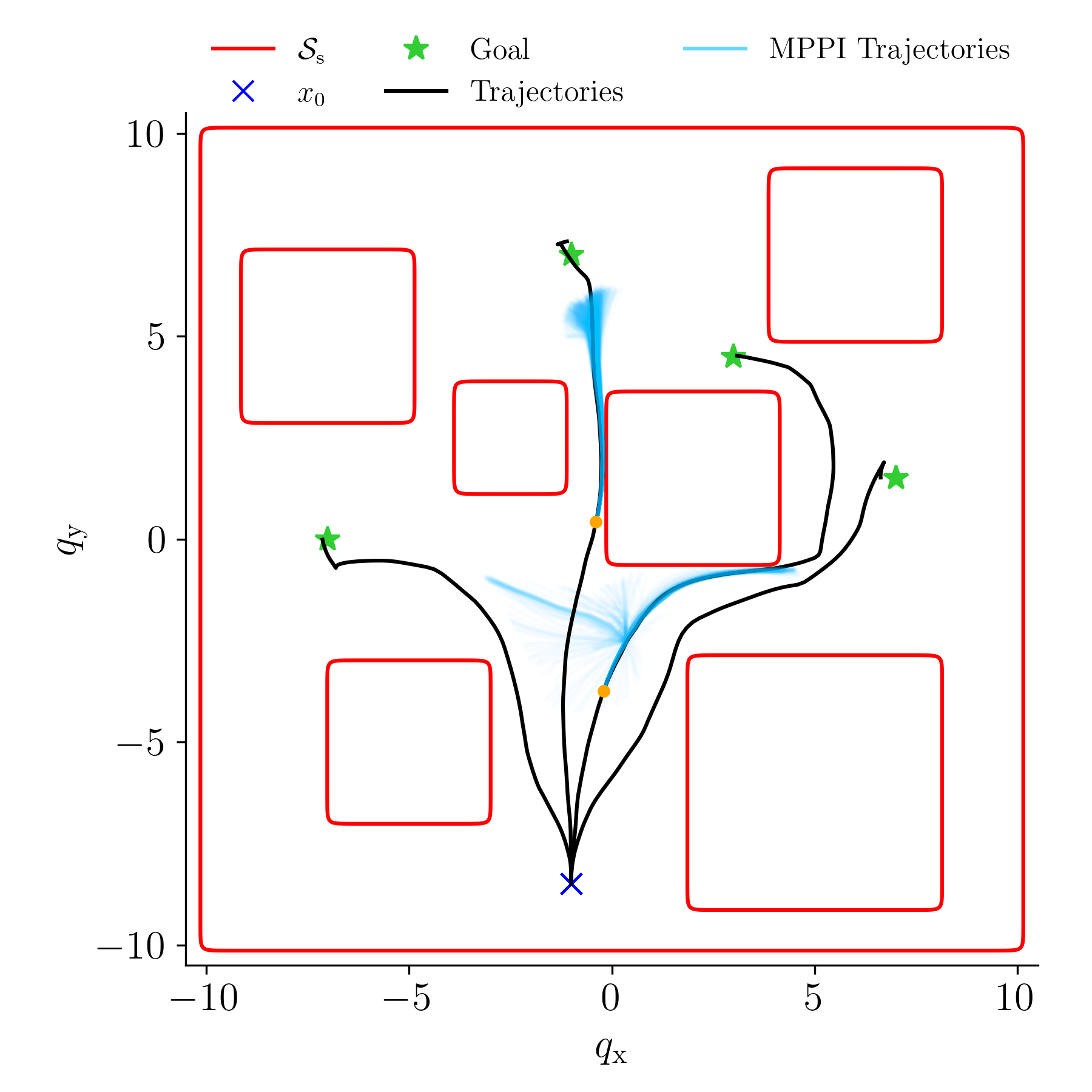}}
\caption{Safe set and closed-loop trajectories using Algorithms \ref{alg:mppi} and \ref{alg:mppi_cbf}}\label{fig:trajs}
\end{figure}

Consider the nonholonomic ground robot modeled by~\eqref{eq:dynamics}, where
\begin{equation*}
    f(x) = \begin{bmatrix}
    \nu \cos{\theta} \\
    \nu \sin{\theta} \\
    0 \\
    0
    \end{bmatrix}, 
    \,
    g(x) = \begin{bmatrix}
    0 & 0\\
    0 & 0\\
    1 & 0 \\
    0 & 1
    \end{bmatrix}, 
    \,
    x = \begin{bmatrix}
    q_\rmx\\
    q_\rmy\\
    \nu\\
    \theta
    \end{bmatrix}, 
    \,
    u = \begin{bmatrix}
    u_1\\
    u_2
    \end{bmatrix}, 
\end{equation*}
and $q \triangleq [ \, q_\rmx \quad q_\rmy \, ]^\top$ is the robot's position in an orthogonal coordinate system, $\nu$ is the speed, and $\theta$ is the direction of the velocity vector (i.e., the angle from $[ \, 1 \quad 0 \, ]^\top$ to $[ \, \dot q_\rmx \quad \dot q_\rmy \, ]^\top$).
Consider the map shown in~\Cref{fig:trajs}, which has 6 obstacles and a wall. 
For $\jmath\in \{1,\ldots,6\}$, the area outside the $\jmath$th obstacle is modeled as the zero-superlevel set of 
\begin{equation}\label{eq:h_obs}
h_\jmath(x) = \left \| \matls a_{\rmx, \jmath} (q_\rmx - b_{\rmx,\jmath})  \\ a_{\rmy, \jmath} (q_\rmy - b_{\rmy, \jmath}) \matrs \right \|_p - c_\jmath,    
\end{equation}
where $b_{\rmx,\jmath}, b_{\rmy,\jmath}, a_{\rmx,\jmath}, a_{\rmy,\jmath}, c_\jmath, p> 0$ specify the location and dimensions of the $\jmath$th obstacle.
Similarly, the area inside the wall is modeled as the zero-superlevel set of
\begin{equation}\label{eq:h_wall}
h_7(x) = c_7 - \left \| \matls a_{\rmx, 7} q_\rmx  \\ a_{\rmy, 7} q_\rmy \matrs \right \|_p,   
\end{equation}
where $a_{\rmx, 7}, a_{\rmy, 7}, c_7 , p> 0$ specify the dimension of the space inside the wall. 
The bounds on speed $v$ are modeled as the zero-superlevel sets of
\begin{equation}
    h_8(x) = 9 - \nu,\qquad 
    h_9(x) = \nu + 1.\label{eq:h_speed.b}
\end{equation}
The safe set $\SSS_\rms$ is given by \eqref{eq:S_s} where $\ell = 9$.
The projection of $\SSS_\rms$ onto the $q_\rmx$--$q_\rmy$ plane is shown in~\Cref{fig:trajs}. 
Note that for all $x\in\SSS_\rms$, the speed satisfies $\nu\in[-1,9]$.
We also note that~\ref{cond:hocbf.a} is satisfied with $d_1 = d_2 = \ldots = d_7 = 2$ and $d_8 = d_9 = 1$.

Let $q_{\rmd} = [\, q_{\rmd,\rmx} \quad q_{\rmd,\rmy}\,]^\rmT \in \BBR^2$ be the goal location, that is, the desired location for $q$.
We consider the cost~\cref{eq:cost_discrete} where $N = 20$, and
\begin{gather*}
\phi(x) = 2\, (q - q_\rmd)^\top (q - q_\rmd),\\
\psi(x, v) = (q - q_\rmd)^\top (q - q_\rmd) + 0.05\,v^\top v,
\end{gather*}

We implement Algorithms \ref{alg:mppi} and \ref{alg:mppi_cbf} with $\alpha_{1,0}(h) = \ldots= \alpha_{6,0}(h) = 2.5h$, $\alpha_{7,0}(h)=h$, and $\alpha(h) = 0.5h$, $\rho = 20$, $\gamma = 10^{24}$, $\Sigma = \matls 1.33 &  0 \\ 0 & 0.33 \matrs$.
$\lambda = 1.0$, $K=1000$, $T_\rms= 0.1\,\rms$, $\delta t = 0.05\,\rms$.

\Cref{fig:trajs} shows 4 closed-loop trajectories for
$x_0=[\,-1\quad -8.5\quad 0\quad {\pi}/{2}\,]^\rmT$ with 4 different goal locations $q_{\rmd}=[\,3\quad 4.5\,]^\rmT$, $q_{\rmd}=[\,-7\quad 0\,]^\rmT$, $q_{\rmd}=[\,7\quad 1.5\,]^\rmT$, and $q_{\rmd}=[\,-1\quad 7\,]^\rmT$. 
In all cases, the robot position converges to the goal location while satisfying safety constraints.

\Cref{fig:trajs} also illustrates the MPPI planning trajectories for two specific scenarios: the goal location $q_{\rmd}=[\,3 \quad 4.5\,]^\rmT$ at time $t = 3\,\rms$ and the goal location $q_{\rmd}=[\,-1 \quad 7\,]^\rmT$ at time $t = 4\, \rms$. These planning trajectories demonstrate a crucial feature of our method: all trajectories generated by Algorithm~\ref{alg:mppi} are guaranteed to be safe. This visual representation demonstrates that our MPPI planner explores only within the safe state space. By generating exclusively safe trajectories, our method significantly improves sample efficiency, a key challenge in MPPI. This approach eliminates the need to evaluate or filter out unsafe trajectories, focusing computational resources on optimizing performance within the safe space. Consequently, our method enhances both computational efficiency and safety assurance in the control process.

\begin{figure}[t!]
\center{\includegraphics[width=0.49\textwidth,clip=true,trim= 0in 0in 0in 0.in] {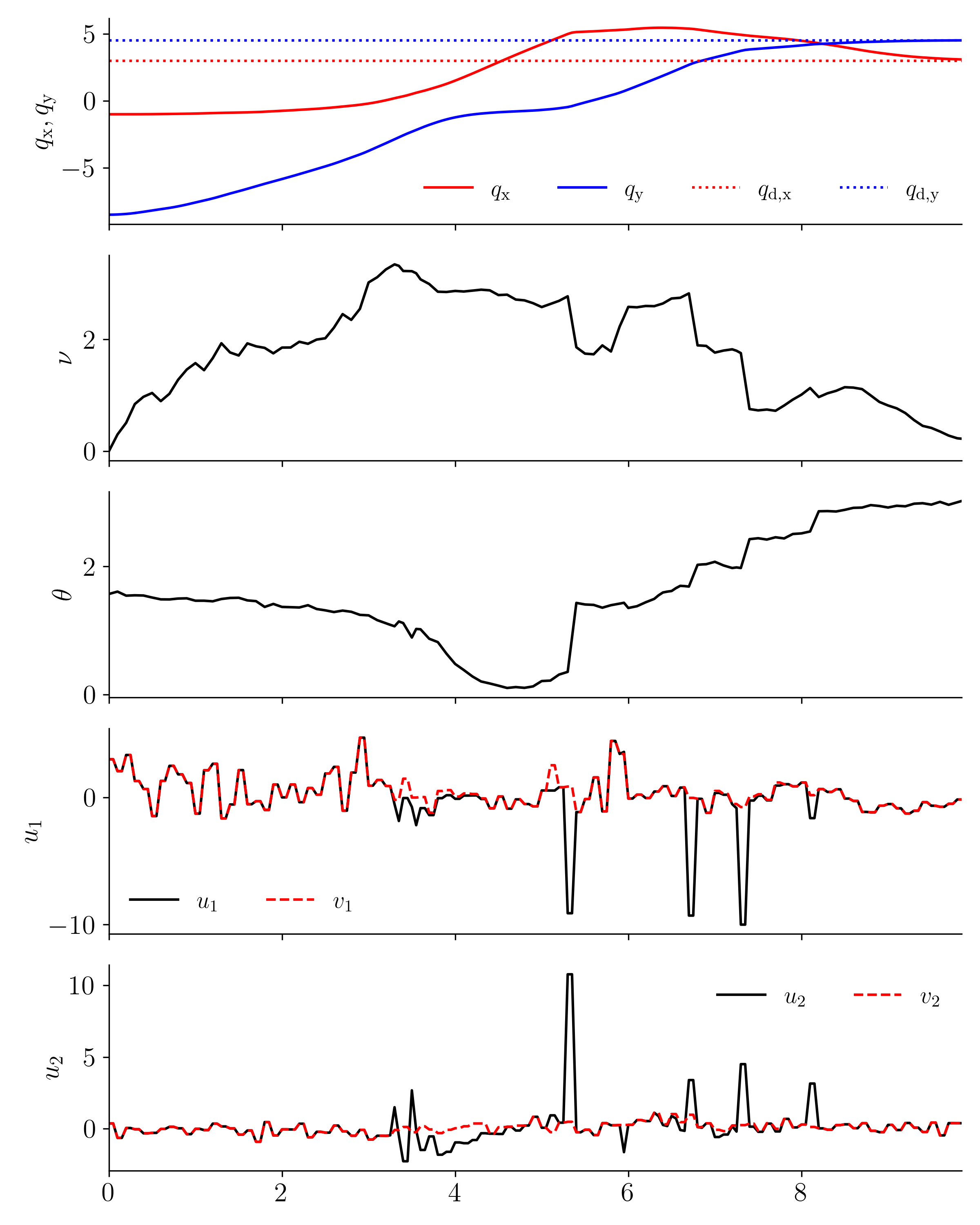}}
\caption{$q_\rmx$, $q_\rmy$, $\nu$, $\theta$, $u$, $v$ for $q_{\rmd}=[\,3 \quad 4.5\,]^\top$.}\label{fig:states}
\end{figure}

\begin{figure}[t!]
\center{\includegraphics[width=0.49\textwidth,clip=true,trim= 0.in 0in 0in 0in] {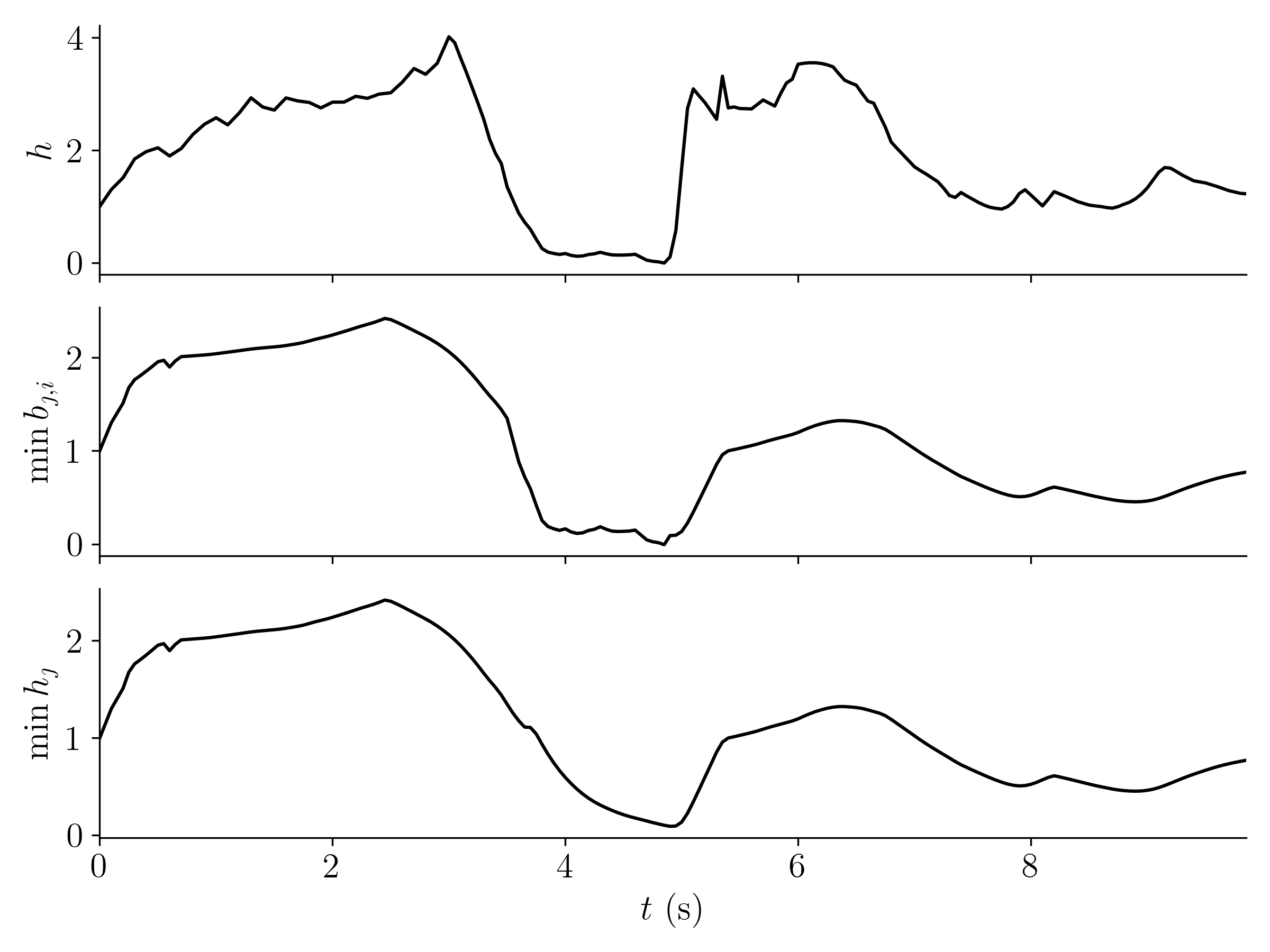}}
\caption{$h$, $\min b_{j,i}$, and $\min h_j$ for $q_{\rmd}=[\,3 \quad 4.5\,]^\rmT$.} \label{fig:barriers}
\end{figure}

\Cref{fig:states} show the trajectories of the relevant signals for the case where $q_{\rmd}=[\,3 \quad 4.5\,]^\rmT$.
Figure~\ref{fig:states}
shows that $h$ is positive for all time, which implies trajectory remains in $\SSS_\rms$.
\Cref{fig:barriers} shows that $h$, $\min b_{\jmath,i}$ and $\min h_\jmath$ are positive for all time, which implies that $x$ remains in $\SSS_\rms$.

 \bibliographystyle{elsarticle-num} 
 \bibliography{mppi_cbf}
 
\end{document}